\documentclass[aps,prc,twocolumn,amsmath,amssymb,superscriptaddress]{revtex4-1}
\usepackage{graphicx}
\usepackage{hyperref}
\usepackage{color}
\newcommand{\be}{\begin{equation}}
\newcommand{\ee}{\end{equation}}
\newcommand{\beq}{\begin{eqnarray}}
\newcommand{\eeq}{\end{eqnarray}}
\newcommand{\ba}{\begin{array}}
\newcommand{\ea}{\end{array}}

\begin{document}

\title{$J/\psi$-Meson Nucleon Scattering Length from Threshold Photoproduction on Light Nuclei}

\date{\today}

\author{\mbox{Igor~I.~Strakovsky}}
\altaffiliation{Corresponding author: \texttt{igor@gwu.edu}}
\affiliation{Institute for Nuclear Studies, Department of Physics, The
    George Washington University, Washington, DC 20052, USA}

\author{\mbox{William~J.~Briscoe}}
\affiliation{Institute for Nuclear Studies, Department of Physics, The
    George Washington University, Washington, DC 20052, USA}

\author{\mbox{Philipp~Gubler}}
\affiliation{Advanced Science Research Center, Japan Atomic Energy Agency,
    Tokai, Naka, Ibaraki 319-1195, Japan}

\author{\mbox{Jackson~R.~Pybus}}
\affiliation{Los Alamos National Laboratory, Los Alamos, NM 87545, USA}

\author{\mbox{Axel~Schmidt}}
\affiliation{Institute for Nuclear Studies, Department of Physics, The
    George Washington University, Washington, DC 20052, USA}

\author{\mbox{Alexander~Somov}}
\affiliation{Thomas Jefferson National Accelerator Facility, Newport News, 
    VA 23606, USA}

\noaffiliation

\begin{abstract}
The quality of recent SRC/CT Collaboration $J/\psi$ photoproduction data off a $^4$He target from Hall~D at Jefferson Laboratory, combined with the feasibility of measuring the reaction close to the free-nucleon energy threshold, opens the door to using incoherent $J/\psi$ photoproduction to access a variety of interesting physics aspects. An example is an estimate of the $J/\psi~p$ scattering length $|\alpha_{J/\psi~p}|$ on the bound proton obtained using the Vector Meson Dominance model. This value can be compared with that of the free proton from the GlueX 
Collaboration~\cite{Strakovsky:2019bev}. One may then project what would be expected from the SRC/CT Collaboration Experiment E12--25--002, which was recently approved by the JLab PAC~\cite{Pybus:2025rba}. Using a plane-wave theoretical model to generate quasi-data, we find the experiment could achieve a result of $|\alpha_{J/\psi~p}| = 3.08\pm 0.45~\mathrm{mfm}$, an uncertainty competitive with that of the free-proton measurement. A comparison between the two would allow an evaluation of the effects of medium modification in the case of light nuclei.  
\end{abstract}

\maketitle

\section{Introduction}
\label{sec:intro} 
The first threshold $J/\psi$ photoproduction measurements on a proton target by the GlueX Collaboration~\cite{GlueX:2019mkq} have enabled a determination of the $J/\psi$-proton scattering length (SL) using Sakurai’s Vector Meson Dominance (VMD) model~\cite{Sakurai:1960ju}. The $J/\psi$ is a member of the vector meson family and has the same spin ($J$), parity ($P$), and charge conjugation ($C$) quantum numbers as the photon: $J^{PC} = 1^{-~-}$. It is assumed that a real photon can fluctuate into a virtual vector meson ($J/\psi$), which then scatters off a nearby proton. The phenomenological analysis of the GlueX threshold data~\cite{GlueX:2019mkq} resulted in a scattering length $|\alpha_{J/\psi p}| = (3.08\pm 0.55)~\mathrm{mfm}$~\cite{Strakovsky:2019bev}; a value much smaller than the size of a hadron ($\approx1~\mathrm{fm}$). This effect can be understood by applying a hypothesis of the ``young'' meson~\cite{Feinberg:1980yu}. 
Due to the small size difference between the ``young'' and ``old'' $J/\psi$, the measured and predicted SL values are very small. $J/\psi$ is created by the photon at the threshold; then, most probably, $J/\psi$ is not fully formed, and its radius is smaller than that of normal (``old'') $J/\psi$~\cite{Feinberg:1980yu}.  Therefore, a stronger suppression of the $J/\psi$-$p$ interaction is 
observed~\cite{Strakovsky:2021vyk}. In a recent study, the effect of the VMD assumption was examined within the formalism of Dyson-Schwinger equations, which can be considered an alternative interpretation of the ``young age'' effect in more formal language~\cite{Xu:2021mju}.

A similar effect may be at play in the case of the $\phi$-nucleon SL, for which recent photoproduction~\cite{Strakovsky:2020uqs} and 
femtoscopy~\cite{ALICE:2021cpv} show a significant disagreement. Theoretical computations~\cite{Lyu:2022imf, Abreu:2024qqo, Feijoo:2024bvn} of this quantity likewise exhibit a wide spread, suggesting that the $\phi$ 
nucleon interaction is, at the moment, still far from being properly understood. 
An alternative constraint can be obtained from the $\phi$ meson mass shift and decay width in nuclear matter, which, within the linear-density approximation, is related to the real and imaginary parts of the $\phi$-nucleon 
SL~\cite{Gubler:2024ovg}. The mass shift of $-3.4\,\%$ reported by the KEK--E325 experiment~\cite{KEK-PS-E325:2005wbm}, for instance, corresponds to an attractive real part of the $\phi$ nucleon SL, with a magnitude of about half that obtained by the ALICE Collaboration. A recent reanalysis of the same data in Ref.~\cite{KEK-PSE325:2025fms}, however, led to a reduced mass shift with significantly larger uncertainties, suggesting that it is still difficult to draw any definite conclusions from the presently available data.

The evaluation of $\gamma A \to J/\psi pX$ cross sections of light nuclei is challenging. The first results on $^4$He were recently reported by the Short-Range Correlations/Color-Transparency (SRC/CT) Collaboration at Jefferson Lab~\cite{Pybus:2024ifi}. Unfortunately, the low statistics do not allow for a sufficiently precise determination of the total near-threshold cross sections to be used as input to determine $|\alpha_{J/\psi~p}|$ on the bound proton. However, a higher-statistics follow-up experiment on $^4$He is planned~\cite{Pybus:2025rba}. Here, we generate quasi-data based on the anticipated statistical precision of this follow-up measurement to assess the precision that can be achieved in extracting $|\alpha_{J/\psi~p}|$.  

\section{Theoretical Developments}
Theoretically, the $J/\psi$-N SL in vacuum has been 
studied in QCD sum rules, effective hadronic models, and lattice QCD approaches, while 
finite density effects on its value have remained largely unexplored. 
In QCD sum rule studies~\cite{Hayashigaki:1998ey, Klingl:1998sr, Kim:2000kj, Yeo:2025ufo}, the SL 
can be extracted either by directly computing the forward $J/\psi$-N scattering amplitude~\cite{Hayashigaki:1998ey} 
or via the $J/\psi$ mass shift in dense matter 
$\Delta m_{J/\psi}(\rho)$~\cite{Klingl:1998sr, Kim:2000kj, Yeo:2025ufo}, which, 
within the linear density approximation, can be related to the 
real part of the SL by 
\begin{equation}
    \mathrm{Re} \alpha_{J/\psi~p} = - \frac{\mu \Delta m_{J/\psi}(\rho)}{2\pi \rho} \>,
    \label{eq:mass_shift}
\end{equation} 
where $\mu$ is the reduced mass of the $J/\psi$-N system. 
In these works, the SL is related mostly to expectation values of gluonic 
operators of a one-nucleon state, demonstrating its usefulness as a probe of the gluonic 
structure of the nucleon, and is in line with an early prediction of a nuclear-bound quarkonium state due 
to a ``QCD van der Waals interaction'' caused by multiple gluon 
exchanges~\cite{Brodsky:1989jd, Brodsky:1997gh} 
and the QCD multipole expansion~\cite{Sibirtsev:2005ex} 
(note, however, the recent Ref.~\cite{Yeo:2025ufo}, in which the relevance of quark-type operators is discussed). 
The above QCD sum rule studies obtain negative mass shifts of the order of $5~\mathrm{MeV}$ at normal nuclear matter density $\rho_0$, which translates 
into SL values of around $\mathrm{Re} \alpha_{J/\psi~p} \sim 0.1~\mathrm{fm}$.

In one example of an effective hadronic model approach~\cite{Krein:2010vp}, the $D$ and $D^{\ast}$ loop contributions to 
the $J/\psi$ self energy are computed with density dependent $D$ and $D^{\ast}$ masses obtained from the quark-meson 
coupling model~\cite{Guichon:1987jp}. This generates an attractive $J/\psi$-N interaction and density-dependent pole masses, 
shifted by about $-20~\mathrm{MeV}$ at $\rho_0$, leading to an SL value of $\mathrm{Re} \alpha_{J/\psi~p} \sim 0.4$ fm.
A more rigorous Effective Field Theory treatment, including a power counting scheme, was introduced in Ref.~\cite{TarrusCastella:2018php}. 
However, more constraints on the different low-energy constants will be essential for this theory to demonstrate real 
predictive power. 
Furthermore, attempts to compare various hadronic models with the 
GlueX photoproduction data of Refs.~\cite{GlueX:2019mkq, GlueX:2023pev} have been made~\cite{Du:2020bqj, JointPhysicsAnalysisCenter:2023qgg}. 
The obtained conclusions w.r.t. the SL, however, exhibit a large spread, ranging from a few mfm~\cite{Du:2020bqj} to values 
of the order of $1~\mathrm{fm}$~\cite{JointPhysicsAnalysisCenter:2023qgg}, suggesting that more complete and precise data will 
be needed to constrain the model parameters and draw any definitive conclusions regarding the value of the $J/\psi$-N SL. 

Finally, Lattice QCD has made progress towards a more quantitative understanding of the $J/\psi$-N interaction. 
While first attempts had to rely on the quenched approximation with rather heavy pion masses~\cite{Yokokawa:2006td}, 
the most recent calculations use $N_f = 2$~\cite{Skerbis:2018lew} or $N_f = 2 + 1$~\cite{Lyu:2024ttm} active quark flavors and a pion 
mass approaching the physical point; however, they yield somewhat contradictory results.  
The calculations in Ref.~\cite{Skerbis:2018lew} of energy levels of the $J/\psi$-N system in a finite box found  
results consistent with almost non-interacting $J/\psi$ and N, corresponding to a vanishing (or very small) SL. 
On the other hand, in Ref.~\cite{Lyu:2024ttm} the $J/\psi$-N potential was calculated using the method proposed by the 
HAL~QCD Collaboration~\cite{Aoki:2012tk}, which 
turns out to be attractive, with a long-range part consistent with a two-pion exchange mechanism. 
The SLs corresponding to this potential are obtained as $0.30(2)^{+0}_{-2}~\mathrm{fm}$ and $0.38(4)^{+0}_{-3}~\mathrm{fm}$ for
the spin 3/2 and 1/2 channels, respectively. 

With all these widely scattered results, both more stringent experimental constraints and a better theoretical understanding of 
the $J/\psi$-N interaction and the $J/\psi$ photoproduction mechanism are clearly needed to obtain a complete picture of the 
physics governing these phenomena.  

\section{Quasi-data for \texorpdfstring{$J/\psi$}{J/psi} Photoproduction at the Threshold on Light Nuclei}
\label{sec:quasi_data}
The underlying cross section model for the quasi-data is the same as that deployed in~\cite{Pybus:2024ifi, Pybus:2025rba}. The model assumes the photoproduction of a $J/\psi$ from a bound proton in the Plane-Wave Impulse Approximation (PWIA), whereby the $J/\psi$ production cross section is factorized from a spectral function that defines the probability that a bound proton has an initial momentum $\vec{p}_i$ and initial energy $\varepsilon_i$. The cross section can be written:
\begin{equation}
    \frac{d\sigma}{dtd\varepsilon_id^3\vec{p}_i} = \frac{p_{\gamma,\mu} p_i^\mu}{\varepsilon_\gamma \varepsilon_i}\cdot \frac{d\sigma}{dt}_{\gamma p \rightarrow J/\psi p} \cdot S(\vec{p}_i,\varepsilon_i)
    \>,
    \label{eq:csA}
\end{equation}
where $p_{\gamma,\mu}=(\varepsilon_\gamma, \vec{p}_\gamma)$ is the four-momentum of the incoming photon beam, $p_{i,\mu} = (\varepsilon_i, \vec{p}_i)$ is the initial four-momentum of the bound proton, $t$ is the Mandelstam variable defined by $t\equiv (p_{\gamma,\mu} - p_{J/\psi,\mu})^2$, where $p_{J/\psi,\mu}$ is the four-momentum of the final state $J/\psi$, $\frac{d\sigma}{dt}_{\gamma p \rightarrow J/\psi p}$ is the $J/\psi$ photoproduction cross section from a single proton, and $S(\vec{p}_i,\varepsilon_i)$ is the spectral function.

Naively, it seems that the ``nuclear effects'' should be rather small for a $^4$He target. The bound state properties of nucleons inside $^4$He are already taken into account by the use of the spectral function in Eq.~(\ref{eq:csA}). There could be some effect from a slight mass change of the $J/\psi$ in a dense medium, but one guesses that this is a small effect that might even be hard to estimate, as there is not really ``dense matter'' in $^4$He, but rather just four nucleons.

Following \cite{Xu:2020,Hatta:2020}, we parameterize $\frac{d\sigma}{dt}_{\gamma p \rightarrow J/\psi p}$ using the form:
\begin{equation}
    \frac{d\sigma}{dt}_{\gamma p \rightarrow J/\psi p} = \sigma_0 (1-\chi)^\beta \frac{F(t) }{\int_{t_\text{min.}}^{t_\text{max.}}F(t)dt} 
    \>,
\label{eq:csp}
\end{equation}
where $\chi$ is an energy fraction parameter defined by 
\begin{equation}
    \chi=\frac{m_{J/\psi}^2 + 2m_{J/\psi} m_p}{s-m_p^2} 
    \>,
    \label{eq:efrac}
\end{equation}
where $m_{J/\psi}$ is the $J/\psi$ mass, $m_p$ is the proton mass, $s$ is the Mandelstam variable defined by $(p_{\gamma,\mu} + p_{i,\mu})^2$, $F(t)$ is a form factor assumed to have a dipole form, \textit{i.e.}, $F(t) = (1-t/\Lambda^2)^{-2}$, $t_\text{min.}$ and $t_{max.}$ are the kinematic limits of $t$ given the photon energy $\varepsilon_\gamma$, and $\sigma_0$, $\beta$, and $\Lambda$ are tunable parameters. The values of these parameters used for this work are given in Table~\ref{tab:cs_params}, and are based on a fit to the latest GlueX data~\cite{GlueX:2023pev}.
\begin{table}[htpb]
    \caption{Parameter values used in the cross section described in 
    Eqs.~(\ref{eq:csA}) and (\ref{eq:csp}).}
    \centering
    \vspace{0.3cm}
    \begin{tabular}{|c|c|}
    \hline
    Parameter  & Value\\
    \hline 
    $\sigma_0$ & $5.9~\mathrm{nb}$   \tabularnewline
    $\beta$    & 1.19                \tabularnewline
    $\Lambda$  & $1.35~\mathrm{GeV}$ \tabularnewline
    \hline
    \end{tabular}
    \label{tab:cs_params}
\end{table}

We used the same approach to the $^4$He spectral function as in ~\cite{Pybus:2024ifi}, divided into a mean-field component calculated using variational Monte Carlo techniques~\cite{RoccoLovato}, and a short-range correlation component estimated using the Generalized Contact Formalism (GCF)~\cite{Weiss:2018tbu, CLAS:2020mom, Pybus:2020itv}.

\begin{figure}[htb!]
\centering
{
    \includegraphics[width=0.45\textwidth,keepaspectratio]{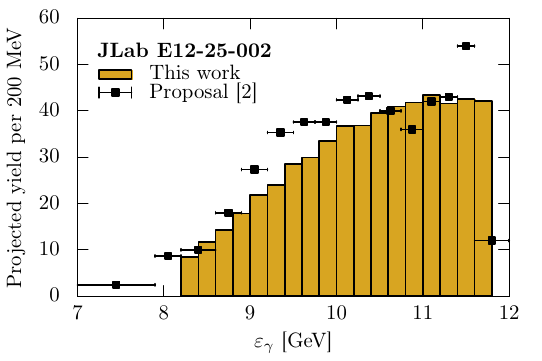}
}
\centerline{\parbox{0.4\textwidth}{
\caption[] {\protect\small
Projected yields for ($\gamma^4\mathrm{He}\to J/\psi pX$) as a function of beam photon energy $\varepsilon_\gamma$. Our estimate, shown as brown bars, is comparable to the estimate provided in the proposal for E12--25--002~\cite{Pybus:2025rba}, which is represented by black points. The horizontal bars indicate the binning of Ref.~\cite{Pybus:2025rba}. }
\label{fig:stat} } }
\end{figure}

To generate quasi-data, an estimate was needed for the tagged photon flux in the experiment. The Hall~D photon beam is produced through the coherent bremsstrahlung process, which leads to a ``coherent peak'' in the produced photon flux at an energy that can be tuned by adjusting the diamond radiator~\cite{GlueX:NIM}. Though the exact flux distribution will depend on the chosen radiator settings and tagger detector placement, we estimated the flux distribution based on the assumptions given in Ref.~\cite{Pybus:2025rba}. Details can be found in Appendix~\ref{app:flux}. Our flux distribution estimate results in a total integrated coherent flux of $5.6\times 10^6~\mathrm{photons/s}$, as well as a total incoherent flux of $4.6\times 10^7~\mathrm{photons/s}$, for a total tagged flux of $5.2\times 10^7~\mathrm{photons/s}$. We assume an experimental duration, $t_\text{exp.}$ of $80~\mathrm{days}$ of full flux on a $30~\mathrm{cm}$ liquid $^4$He target (total thickness, $n_\text{tar.}$, of $5.68\times 10^{23}~\mathrm{nuclei/cm^2}$). We also assume that the reconstructed yields will be reduced both by nuclear attenuation---the final-state particles can be absorbed or re-scattered by spectator nucleons before emerging from the nucleus---and by experimental efficiency. Based on the results of the Glauber model calculation performed in Ref.~\cite{Pybus:2024ifi}, we assume a conservative nuclear transparency factor, $T$, of 60\%, independent of kinematics. We assume a reconstruction efficiency, $\epsilon$ of 15\%, also based on Ref.~\cite{Pybus:2024ifi}. Our estimate of the yield in a given bin can be written:
\begin{equation}
    Y_\text{bin} = b_{e^+e^-}\cdot \epsilon \cdot T \cdot n_\text{tar.} \cdot t_\text{exp.} \cdot \int_\text{bin} \Phi(\varepsilon_\gamma) \sigma(\varepsilon_\gamma) d\varepsilon_\gamma 
    \>,
\end{equation}
where $b_{e^+e^-}=6\%$ is the branching fraction for $J/\psi$ decay into $e^+e^-$~\cite{ParticleDataGroup:2024cfk} and $\Phi(\varepsilon_\gamma)$ is the photon flux distribution. Our estimated yields for E12--25--002, in 18 bins of $\varepsilon_\gamma$ are shown in Fig.~\ref{fig:stat}, and are similar to the estimates made in Ref.~\cite{Pybus:2025rba}. We do not consider subthreshold quasi-data in the analysis.

To estimate the uncertainty per bin, we combine the Poisson statistical uncertainty with 14\% systematic uncertainty, as suggested in Ref.~\cite{Pybus:2025rba}, 
covering the effects of $\pi^+\pi^-$ contamination, background subtraction, cut-dependence, uncertainty on the Bethe-Heitler reference reaction, and efficiency. 

\begin{figure}[htb!]
\centering
{
    \includegraphics[width=0.45\textwidth,keepaspectratio]{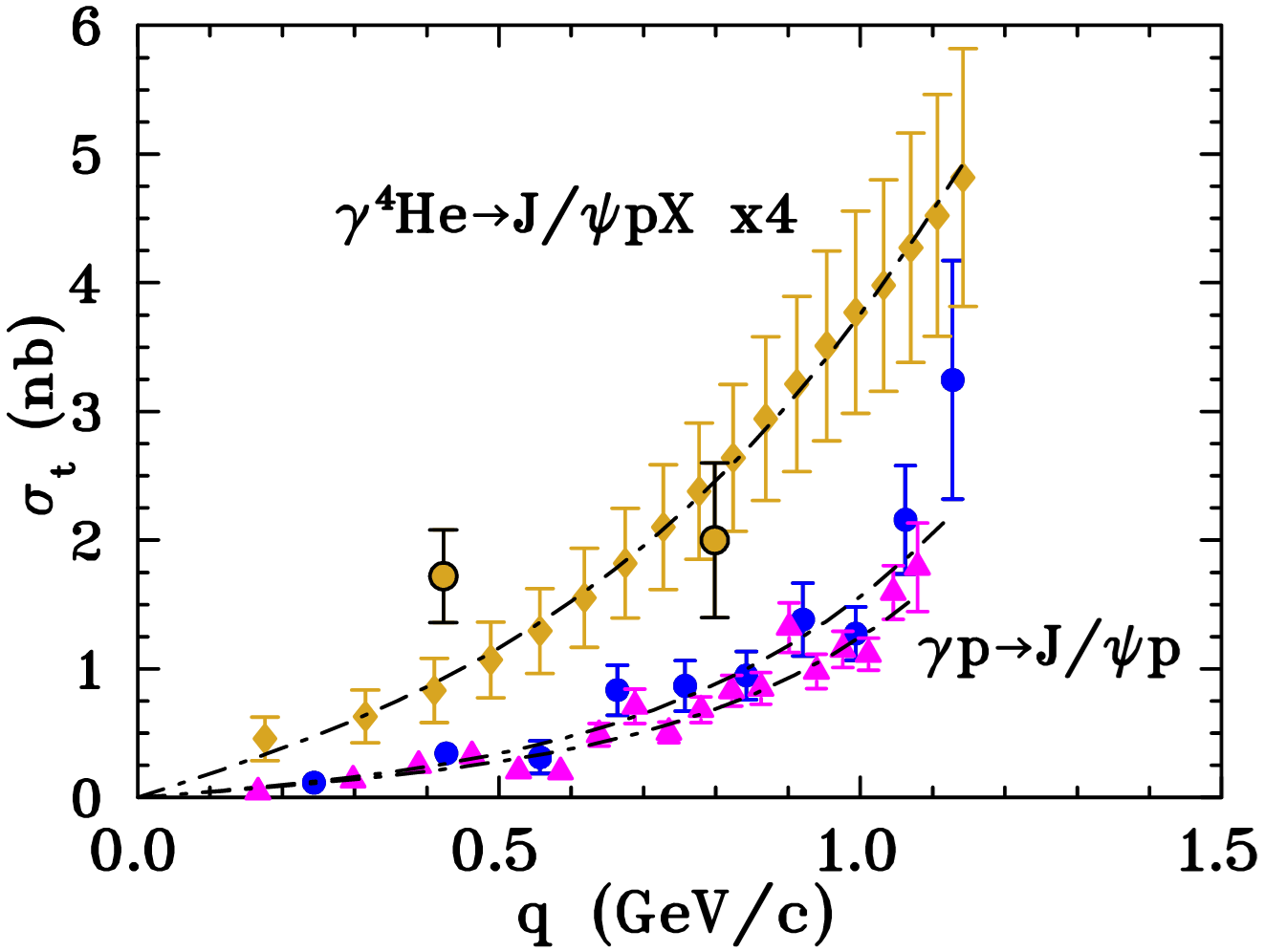}
} 
\centerline{\parbox{0.4\textwidth}{
\caption[] {\protect\small
The total $\gamma p \to J/\psi p$ cross section $\sigma_t$ 
as a function of the $J/\psi$ CM momentum includes 
measurements of the GlueX Collaboration with small statistics  (blue filled circles, $469\pm 22~\mathrm{events}$)~\cite{GlueX:2019mkq,Strakovsky:2019bev}  and  with high statistics (magenta filled triangles, $2270\pm 
58~\mathrm{ events}$)~\cite{GlueX:2023pev}. Quasi-data for the total $\gamma ^4\mathrm{He} \to J/\psi~X$ cross section obtained from GCP, are shown as brown diamonds. Experimental results from the SRC/CT Collaboration are indicated by filled circles with black rings~\cite{Pybus:2024ifi}.
The black dash-dotted curves represent polynomial fits to the the proton data (Eq.~(\ref{eq:eq2})) and to the light nucleus ($^4$He) data (Eq.~(\ref{eq:eq3})).
} 
\label{fig:fig2} } }
\end{figure}

The $J/\psi$ cross section and corresponding uncertainty obtained using simulated quasi-data is shown in Fig.~\ref{fig:fig2} as a function of the 
center-of-mass (CM) momentum $q$ of the $J/\psi$. In computing the CM momentum, we neglected Fermi motion effects; these can be incorporated into an experimental analysis by binning the data directly in $q$ and deconvoluting the nuclear spectral function. The uncertainties associated with this procedure are expected to be significantly smaller than those arising from the determination of the total cross section. The scattering length and its associated uncertainties are then extracted from a fit to these cross sections, as described in the next section.

\section{\texorpdfstring{$J/\psi$}{J/psi} Nucleon Scattering length: Phenomenological analysis}
For the evaluation of the \textit{absolute value} of $J/\psi~p$ SL, we apply the VMD approach that links the near-threshold photoproduction total cross section of $\gamma p \to J/\psi p$ and elastic $J/\psi p \to J/\psi p$~\cite{Strakovsky:2014wja, Strakovsky:2021vyk}.
To avoid theoretical uncertainties, we do not (i) determine the sign of SL, (ii) separate the Re and Im parts of SL, nor (iii) extract total angular momentum 1/2 and 3/2 contributions for the $J/\psi~N$ system. The limitations of VMD are listed in Refs.~\cite{Xu:2021mju, Kopeliovich:2017jpy, Boreskov:1976dj}.

Traditionally, the total cross section, $\sigma_t$, for near-threshold binary inelastic reactions  
\begin{equation}
    a~b\to c~d
\label{eq:eq0}
\end{equation}
with 
\begin{equation}
    m_a + M_b < m_c + M_d 
\label{eq:eq1}
\end{equation}
is expressed as a series in \textit{odd} powers of the CM momentum $q$ of the $J/\psi$ meson 
\begin{equation}
    \sigma_t = a~q + b~q^3 + c~q^5
    \>,
\label{eq:eq2}
\end{equation}
where the linear term reflects contributions of two independent $S$-wave amplitudes with a total spin of 1/2 and/or 3/2. Contributions to the cubic term originate from both $P$-wave amplitudes and the energy ($W$) dependence of $S$-wave amplitudes, and the fifth-order term arises from $D$-wave amplitudes as well as from the $W$ dependencies of the $S$- and $P$-wave amplitudes.

For a nuclear target, Eq.~(\ref{eq:eq2}) retains the same structure, but the fit is performed for $\sigma_t/Z$ instead of just $\sigma_t$:
\begin{equation}
    \sigma_t/Z = a~q + b~q^3 + c~q^5
    \>,
\label{eq:eq3}
\end{equation}
where $Z$ is the charge of the nuclear target (in the case of $^4$He, $Z = 2$). In this analysis, we determine $q$ from $\varepsilon_\gamma$ assuming a stationary nucleon target, \textit{i.e.}, neglecting Fermi motion effects.

The absolute value of SL is determined by the interplay between strong (hadronic) and electromagnetic dynamics.
\begin{equation}
    |\alpha_{J/\psi p}| = h_{J/\psi p}~B_{J/\psi}
    \>,
\label{eq:eq4}
\end{equation}
where the hadronic component is given by
\begin{equation}
    h_{J/\psi p} = \sqrt{a}
    \>,
\label{eq:eq5}
\end{equation}
and the parameter $a$ is the linear term obtained from the best fit to the total cross section  $\sigma_t$ of the V photoproduction (Eqs.~(\ref{eq:eq2}) and (\ref{eq:eq3})).
The purely electromagnetic, VMD-motivated kinematic factor is
\begin{equation}
    B_{J/\psi}^2 = \frac{\alpha~m_{J/\psi}~k}{12~\pi ~\Gamma(J/\psi \to e^+e^-)}
    \>,
\label{eq:eq6}
\end{equation}
where $\alpha$ is the fine-structure constant,
$m_{J/\psi}$ is the mass of $J/\psi$ ($m_{J/\psi} =3096.900\pm 
0.006~\mathrm{MeV}$)~\cite{ParticleDataGroup:2024cfk},
$k$ is the photon CM momentum, and
$\Gamma(J/\psi \to e^+e^-)$ is the $J/\psi$ partial decay width to 
$e^+e^-$, $\Gamma(J/\psi \to e^+e^-) = 5.53\pm 0.10~\mathrm{keV}$~\cite{ParticleDataGroup:2024cfk}.

\vspace{5cm}
\begin{table}[htb!]
\centering \protect\caption{
The $J/\psi~p$ scattering lengths extracted from the GlueX measurements on a free proton and 
from the SRC/CT experiment on a ${}^4{\rm He}$ target. The second column lists the minimal 
momentum $q_{min}$ for $J/\psi$ accessible in the photoproduction experiments. The third 
column gives the linear coefficient  $a$ obtained from fits to the $J/\psi$ production cross 
sections using Eqs.~(\ref{eq:eq2}) and (\ref{eq:eq3}). The fourth column presents the corresponding scattering lengths, derived from the parameter $a$ via the phenomenological approach.
}
\vspace{2mm}
{%
\begin{tabular}{|c|c|c|c|}
\hline
Target   & $q_{min}$ & $a$   & $|\alpha_{J/\psi p}|$ \tabularnewline
         & (MeV/c)   & ($\times 10^{-6}~\mathrm{\mu b/(MeV/c)}$) & (mfm) \tabularnewline
\hline
$^1$H    & 244~\cite{GlueX:2019mkq}      & 0.46$\pm$0.16~\cite{Strakovsky:2019bev}  & 3.08$\pm$0.55~\cite{Strakovsky:2019bev}
\tabularnewline
$^1$H    & 166~\cite{GlueX:2023pev}      & 0.43$\pm$0.10 & 2.98$\pm$0.25 \tabularnewline
$^4$He   & 176                           & 0.46$\pm$0.13 & 3.08$\pm$0.45 
\tabularnewline
\hline
\end{tabular}} \label{tbl:tab1}
\end{table}

\section{Measurements with the GlueX Detector}
The photoproduction of the $J/\psi$ meson was studied using the 
GlueX detector, which is located in Hall~D at Jefferson Lab. GlueX is a forward 
magnetic spectrometer designed to carry out experiments with a photon 
beam of up to $12~\mathrm{GeV}$. The photon beam is produced via the bremsstrahlung 
process from the primary electron beam delivered by the Continuous Electron 
Beam Accelerator Facility (CEBAF). $J/\psi$ mesons are reconstructed 
through the decay channel $J/\psi \to e^+e^-$. Charged particles are tracked 
using the central drift chamber, positioned inside the solenoid magnet, 
together with a set of forward planar drift chambers. Lepton identification 
is performed using the forward and barrel calorimeters, which also provide 
pion rejection by applying selection criteria on the ratio $p/E$, where 
$p$ is the particle momentum and $E$ is the energy deposited in the 
calorimeters. Timing information from scintillating detectors, the start 
counter surrounding the target, and the forward time-of-flight wall 
further contribute to particle identification.

The first measurement of $J/\psi$ photoproduction on the proton was 
performed by the GlueX experiment using the small data sets collected 
in 2016 and 2017. This initial analysis yielded $469\pm22$ reconstructed 
$J/\psi$ events~\cite{GlueX:2019mkq}. A subsequently conducted, significantly larger data sample resulted 
in $2270\pm58$ reconstructed candidates, collected from 2016 to 2018, with the analysis reported in 
Ref.~\cite{GlueX:2023pev}. The $J/\psi$ photoproduction cross section as a function of the 
$J/\psi$ CM momentum is shown in Fig.~\ref{fig:fig2}. The 
scattering lengths extracted from the data are listed in Table~\ref{tbl:tab1} 
and demonstrate good agreement between the higher-statistics analysis  
and the initial publication.

Following GlueX, a short Jefferson Lab experiment was performed to study 
short-range correlations using photon beams. Data were collected on 
deuterium, helium, and carbon targets. This experiment obtained a small 
data sample consisting of $120\pm15$ reconstructed $J/\psi$ events and 
reported the first measurements of near- and sub-threshold $J/\psi$ production 
from nuclear targets~\cite{Pybus:2024ifi}.

Recently, the SRC/CT Collaboration proposed a new experiment (E12--25--002), \textit{``Threshold $J/\psi$ photoproduction as a probe of nuclear gluon 
structure''}~\cite{Pybus:2025rba}, which was approved by the Jefferson Lab 
Program Advisory Committee in 2025 to run for $85~\mathrm{days}$ in Hall~D using the 
GlueX spectrometer. The experiment plans to collect $80~\mathrm{days}$ production 
data on a liquid $^4$He target. 
The experiment will feature an upgraded forward electromagnetic calorimeter 
based on high-resolution, high-granularity lead-tungstate crystals, as well 
as a new  GEM-TRD detector to reduce background from $e/\pi$ 
misidentification.  With a longer run duration and better optimization in
 experimental settings, the experiment aims to increase the 
statistics for $J/\psi$ production from bound protons by a factor of 
approximately 10 compared to the earlier SRC experiment. 
This will enable precision measurements of the incoherent $J/\psi$ 
photoproduction cross section at threshold, providing the first stringent 
constraints on nuclear modifications of gluon structure or other exotic 
effects. Based on our analysis of the quasi-data described in 
Sec.~\ref{sec:quasi_data}, 
we estimate that the total cross section data from the experiment 
will allow for a competitive determination of the $J/\psi$-p scattering length, 
with an uncertainty on the order of $0.45~\mathrm{mfm}$. The total cross section for 
the quasi-data is shown in Fig.~\ref{fig:fig2}, and the corresponding scattering length is listed in Table~\ref{tbl:tab1}. \\

The results of our phenomenological quasi-data analysis are in good agreement with the recent theoretical study of vector-meson photoproduction from threshold to high energies reported by Roberts \textit{et al.}. A good description of the data (without fitting) is achieved primarily by the explicit implementation of the upper quark loop shown in Fig.~1 of Ref.~\cite{Tang:2025qqe}. This approach yields a scattering length of $|\alpha_{J/\psi p}| = 3.08~\mathrm{mfm}$.

An alternative approach for deciphering the mechanism of near-threshold $J/\psi$ photoproduction yielded similar SL~\cite{Du:2020bqj}. $|\alpha^{J=1/2}_{J/\psi p}| = 0.2\cdots 3.1~\mathrm{mfm}$ and $|\alpha^{J=3/2}_{J/\psi p}| = 0.2\cdots 3.0~\mathrm{mfm}$, where $J$ denotes the total angular momentum of the $J/\psi$-nucleon system.

\section{Summary}
The GlueX detector in Hall~D at Jefferson Lab provides a unique 
opportunity to study the photoproduction of $J/\psi$ mesons over a 
broad kinematic range near threshold. The scattering length for $J/\psi$-p  
production on a free proton has already been extracted from measurements 
performed by the GlueX collaboration. The SRC/CT collaboration carried out 
the first measurements of the $J/\psi$ photoproduction cross section on light nuclear targets using a small sample of $J/\psi$ mesons. A recently approved experiment will increase the available statistics by roughly an order of magnitude. We evaluated the feasibility of extracting the scattering length 
for this new experiment using quasi-data. Within the VMD and the plane-wave approximation framework, we 
demonstrated that the upcoming measurement will enable the first determination of the scattering length on a bound proton.



\section{Acknowledgments}
This work was supported in part by the U.S.~Department of Energy, Office of Science, Office of Nuclear Physics, under Awards No. DE--SC0016583 and No. DE--AC05--06OR23177.
Research presented in this article was supported by the Laboratory Directed Research and Development program of Los Alamos National Laboratory (J.R.P.) under project number 20240866PRD2.
P.G. is supported by JSPS KAKENHI Grant Numbers JP22H00122 and JP25H00400. 

\appendix
\section{Flux Model}
\label{app:flux}
To estimate yields for the future Jefferson Lab Experiment E12--25--002~\cite{Pybus:2025rba}, we made 
a rough estimate of a flux distribution, $\Phi(\varepsilon_\gamma)$, given by:
\begin{equation}
    \Phi(\varepsilon_\gamma) = \Phi_\text{coh.}(\varepsilon_\gamma) + \Phi_\text{incoh.}~(\varepsilon_\gamma)
\end{equation}
Switch
\begin{align}
    \Phi_\text{incoh.}(\varepsilon_\gamma) 
    & = \Phi_a + (\Phi_b-\Phi_a)~\frac{\varepsilon_\gamma - \varepsilon_a }{\varepsilon_b - \varepsilon_a} & \varepsilon_a < \varepsilon_\gamma < \varepsilon_b,\\
    \Phi_\text{coh.}(\varepsilon_\gamma) 
    & = \Phi_c~\frac{\varepsilon_\gamma - \varepsilon_d }{\varepsilon_c - \varepsilon_d}  
    & \varepsilon_d < \varepsilon_\gamma < \varepsilon_c 
    \>,
\end{align}
with $\varepsilon_a = 6.0~\mathrm{GeV}$ and 
$\varepsilon_b = 10.8~\mathrm{GeV}$ describing the tagger energy range, 
$\varepsilon_d = 7.5~\mathrm{GeV}$ describing the start of the coherent peak, 
$\varepsilon_c = 8.0~\mathrm{GeV}$ describing the coherent edge, and flux constants $\Phi_a = 1.2\times 10^7~\mathrm{photons/GeV/s}$, $\Phi_b =
4\times 10^6~\mathrm{photons/GeV/s}$, and $\Phi_c = 2.24\times 
10^7~\mathrm{ photons/GeV/s}$. Under this assumption, the total integrated coherent flux is $5.6\times 10^6~\mathrm{photons/s}$, while the total incoherent flux is $4.64\times 10^7~\mathrm{photons/s}$, resulting in a total tagged flux of $5.2\times 10^7~\mathrm{photons/s}$. We estimate that this roughly corresponds to the tagged flux produced by a $12~\mathrm{GeV}$, $160~\mathrm{nA}$ primary electron beam incident on a $4\times 10^{-4}$ radiation length diamond radiator. The precise tagged-flux distribution in the experiment will depend on these experimental settings and on the placement of the various elements of the tagger detector. 


\end{document}